\documentclass[sigconf]{acmart}

\usepackage{booktabs} 
\usepackage{amsmath}
\DeclareMathOperator*{\argmin}{arg\,min}

\usepackage{amscd,amsfonts,amsbsy,rotating}
\usepackage{balance}
\usepackage{graphicx}
\usepackage{epsfig,epstopdf}
\usepackage{subfigure}
\usepackage{multirow}
\usepackage{color,xcolor}
\usepackage{url}
\usepackage{latexsym,bm}
\usepackage{enumitem,balance,mathtools}
\usepackage{wrapfig}
\usepackage{euscript}
\usepackage{algorithm}
\usepackage{algorithmic}
\usepackage{ifpdf}
\usepackage{diagbox}
\usepackage{caption}
\usepackage{makecell}
\usepackage{subfigure}
\usepackage{bm}
\usepackage{float}
\usepackage{threeparttable}

\newcommand{\bs}{\boldsymbol}

\newcommand{\btheta}{\bs{\theta}}

\AtBeginDocument{%
  \providecommand\BibTeX{{%
    \normalfont B\kern-0.5em{\scshape i\kern-0.25em b}\kern-0.8em\TeX}}}

\copyrightyear{2021} 
\acmYear{2021} 
\setcopyright{acmcopyright}
\acmConference[KDD '21]{Proceedings of the 27th ACM SIGKDD Conference on Knowledge Discovery and Data Mining}{August 14--18, 2021}{Virtual Event, Singapore}
\acmBooktitle{Proceedings of the 27th ACM SIGKDD Conference on Knowledge Discovery and Data Mining (KDD '21), August 14--18, 2021, Virtual Event, Singapore}
\acmPrice{15.00}
\acmDOI{10.1145/3447548.3467216}
\acmISBN{978-1-4503-8332-5/21/08}

\begin{document}
\fancyhead{}
\title{Retrieval \& Interaction Machine for Tabular Data Prediction}

\author{Jiarui Qin$^{\dag}$, Weinan Zhang$^{\dag}$*, Rong Su$^{\ddag}$, Zhirong Liu$^{\ddag}$, Weiwen Liu$^{\ddag}$ \and Ruiming Tang$^{\ddag}$, Xiuqiang He$^{\ddag}$, Yong Yu$^{\dag}$}
\affiliation{
  \institution{$^{\dag}$Shanghai Jiao Tong University, $^{\ddag}$Huawei Noah's Ark Lab\\
    \{qinjr, wnzhang, yyu\}@apex.sjtu.edu.cn, \{surong3, liuzhirong, liuweiwen8, tangruiming, hexiuqiang1\}@huawei.com}
    \country{}
}

\renewcommand{\shortauthors}{J.Qin, et al.}

\begin{abstract}
  Prediction over tabular data is an essential task in many data science applications such as recommender systems, online advertising, medical treatment, etc.
  Tabular data is structured into rows and columns, with each row as a data sample and each column as a feature attribute.
  Both the columns and rows of the tabular data carry useful patterns that could improve the model prediction performance.
  However, most existing models focus on the cross-column patterns yet overlook the cross-row patterns as they deal with single samples independently. 
  In this work, we propose a general learning framework named \textbf{R}etrieval \& \textbf{I}nteraction \textbf{M}achine (RIM) that fully exploits both cross-row and cross-column patterns among tabular data.
  Specifically, RIM first leverages search engine techniques to efficiently retrieve useful rows of the table to assist the label prediction of the target row, then uses feature interaction networks to capture the cross-column patterns among the target row and the retrieved rows so as to make the final label prediction.
  We conduct extensive experiments on 11 datasets of three important tasks, i.e., CTR prediction (classification), top-$n$ recommendation (ranking) and rating prediction (regression). Experimental results show that RIM achieves significant improvements over the state-of-the-art and various baselines, demonstrating the superiority and efficacy of RIM.
  \noindent\let\thefootnote\relax\footnotetext{*~Weinan Zhang is the corresponding author.}
\end{abstract}




\maketitle

\section{Introduction} \label{sec:intro}
\begin{figure*}[t]
    \centering
    \includegraphics[width=1.0\textwidth]{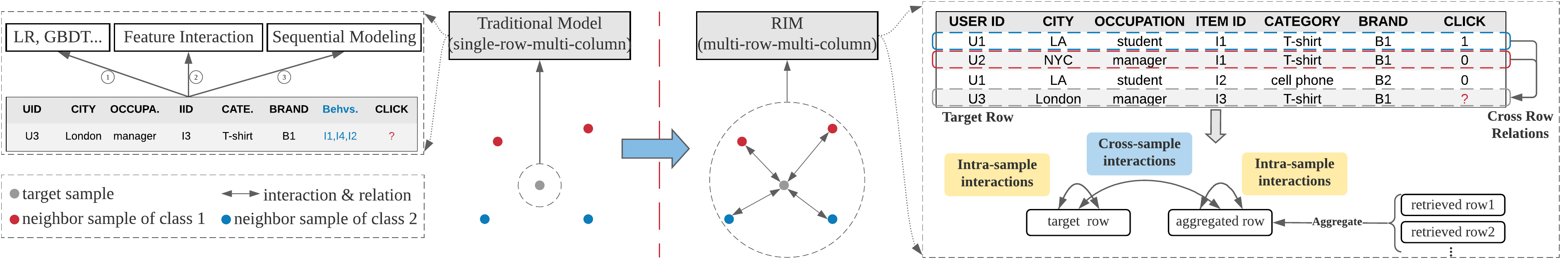}
    \caption{Illustration of RIM and traditional models on tabular data.}
    \label{fig:tabular-rim-illu}
\end{figure*}

Prediction over tabular data is an essential task in various real-world data science applications including click-through rate (CTR) prediction in online advertising \cite{zhou2018deep,zhou2019deep,qu2018product}, rate prediction or item ranking in recommender systems \cite{bobadilla2013recommender}, medical treatment \cite{kononenko2001machine} and fraud detection \cite{bolton2002statistical}, etc.
The majority of collected data of these applications is organized in the form of tables. Each row corresponds to a sample, while each column corresponds to a distinct feature.
It is illustrated in the right part of Figure~\ref{fig:tabular-rim-illu}.


Extracting informative relations or patterns in tabular data is crucial for a learning system to make accurate label prediction. To this end, the potentials of deepening utilization of tabular data, both in rows and in columns, have become more apparent. 
Early models like logistic regression, SVM and tree models directly take as input the features of a row and make predictions. 
Deep learning models like Wide \& Deep \cite{cheng2016wide} or DeepCrossing \cite{shan2016deep} improve the expressive ability by projecting each categorical feature into an embedding and use DNNs to model the tabular data. 
In recent years, feature interaction based models and sequential models are two primary methods for tabular data. 
Feature interaction based models \cite{juan2016field,zhang2016deep,qu2016product,guo2017deepfm,lian2018xdeepfm,qu2018product}, originated from POLY2 \cite{chang2010training} and factorization machine (FM) \cite{rendle2010factorization}, introduce high-order feature combinations and interactions between columns of the tabular data.  
Sequential models, on the other hand, aim to capture patterns in some sequential feature fields. The sequential feature is usually a multi-value feature (such as an item sequence that a user has clicked). These models often incorporate architectures like RNNs, CNNs, Transformer, attention mechanism, or memory networks to learn a dense representation of the sequential feature.
There are two ways that the sequential feature is generated. The common way is using the most recent consecutive behaviors. The other way is retrieving \cite{qin2020user,qi2020search} some relevant behaviors. The retrieval procedure only affects how the sequential feature is generated.
All of the above models could be categorized into the \textit{single-row-multi-column} framework. Because these models only utilize the cross-column patterns of a single row for its label prediction as illustrated in the left part of Figure~\ref{fig:tabular-rim-illu}.

Despite the great success of the existing models that incorporate the cross-column patterns, we argue that it is insufficient to simply input a single row solely.
Inspired by the concept of $k$ nearest neighbor ($k$NN) that makes predictions according to the neighbors, it could be beneficial if we get the relevant samples to assist the inference on the target sample. 
As shown in the right part of Figure~\ref{fig:tabular-rim-illu}, the prediction model should be aware of the existence of the neighbor samples and the interactions between the target sample and its neighbors could be utilized to enhance the prediction.
Both the features and labels of the neighbor samples are strong signals and indicators of the prediction target, which could be used as promising discriminative features to enhance the predictive power.
In the view of the tabular data, the neighbor samples are essentially other rows of a table. Therefore, this learning framework we proposed could be categorized as \textit{multi-row-multi-column} because it not only cares about cross-column patterns in the target row but models the cross-row patterns.

Having realized the vital role of neighbor samples, we focus on exploring how to efficiently obtain the neighbors of a target sample from the entire sample space. 
It is impractical and computationally expensive to calculate the similarity between every pair of samples in a very large sample space. 
As tabular data is highly structured and categorical, we find it reasonable to use search engine techniques to \textbf{retrieve} related rows from the table.
Thus we propose to bridge the $k$NN idea with deep learning models over tabular data by using an elegant retrieval procedure to model the cross-row patterns.

In this work, we propose a general multi-row-multi-column framework called \textbf{R}etrieval \& \textbf{I}nteraction \textbf{M}achine (RIM) that fully exploits cross-row and cross-column patterns among tabular data.
As the name implies, RIM uses a retrieval mechanism to capture the cross-row patterns and utilizes a feature interaction network to model the cross-column patterns at the same time. 
The basic idea of RIM is illustrated in right part of Figure~\ref{fig:tabular-rim-illu}. For the target row whose label needs to be predicted, we firstly retrieve some other rows in the table using search engine techniques. 
Specifically, each row in the table is regarded as a document, and the target row is formulated as a query. Finding the relevant samples could be modeled as a query-document matching problem. Thus we could use various ranking functions such as BM25 \cite{robertson1995okapi} to retrieve the relevant rows. 
After the relevant rows are retrieved, a target-row-aware order-invariant aggregation operation is performed. 
A feature interaction network is then used to capture both the intra-sample feature interactions and the cross-sample feature interactions, as illustrated in the right part of Figure~\ref{fig:tabular-rim-illu}.
In this way, both cross-row and cross-column patterns are captured to enable RIM to make the final prediction.

The contributions of this paper are summarized as follows.
\begin{itemize}[leftmargin=15pt]
    \item This paper proposes RIM, a \textit{multi-row-multi-column} framework which could capture both cross-row and cross-column patterns in tabular data for accurate label prediction. RIM is a generic supervised learning framework over tabular data regardless of the specific tasks.
    \item To our knowledge, RIM is the first framework that introduces a sample-level retrieval module to directly retrieve useful samples from the raw sample space of tabular data, which makes it feasible to obtain relevant rows effectively. The retrieved samples are proved effective in boosting prediction performance.
\end{itemize}

Extensive experiments are conducted on 11 datasets over three different tasks, including CTR prediction (classification), sequential top-$n$ recommendation (ranking), and rating prediction (regression) tasks. These experiments cover both implicit feedback and explicit feedback scenarios. Experimental results show that RIM achieves significant improvements over the state-of-the-art and various baselines, demonstrating its superiority and efficacy.



\section{Preliminaries} \label{sec:preli}
In this section, we formulate the problem and introduce the notations. The tabular dataset is denoted as a table $\mathcal{T}$ with $F$ feature columns $\mathcal{C} = \{\mathcal{C}_i\}_{i=1}^F$ and one label column $\mathcal{Y}$. The feature columns could be features such as "Age", "City", "Occupation", and etc.
Each row of the table $\mathcal{T}$ is one sample which is denoted as $s_z$. Each sample consists of multiple features and a label, denoted as $s_z = (x_z, y_z)$ where $x_z = \{c_i^z\}_{i=1}^F$. Thus the table $\mathcal{T}$ is formulated as the set of $N$ samples as $\mathcal{T} = \{s_z\}_{z=1}^N$. 
In our framework, the entire dataset could be split into three disjoint tables as $\mathcal{T}_{train}$, $\mathcal{T}_{test}$, $\mathcal{T}_{pool}$, where $\mathcal{T}_{train}$ and $\mathcal{T}_{test}$ are normal train and test datasets, while $\mathcal{T}_{pool}$ is the pool from which the historical samples are retrieved. The details of how to split the dataset will be described in Section~\ref{sec:exp-setting}. 

The goal of modeling tabular data is to predict the label based on the feature columns of a specific sample. The traditional framework is formulated as
\begin{equation} \label{eq:old-formulation}
    \hat{y}_z = \mathit{f}(x_z; \btheta),
\end{equation}
where $\mathit{f}$ is the learned function with parameters $\btheta$.
We argue that using the retrieved similar samples could help the prediction performance, so in our framework, the goal is formulated as 
\begin{equation} \label{eq:rim-formulation}
    \hat{y}_z = \mathit{f}(x_z, \mathcal{R}(x_z); \btheta),
\end{equation}
where $\mathcal{R}(x_z)$ represents the retrieved neighbor samples of $s_z$. The retrieval process is non-parametric and $\btheta$ is the parameters of the prediction function.
After getting the predicted value, the parameters are optimized by minimizing the loss function as
\begin{equation}
    \btheta^* = \argmin_{\btheta} \sum_{s_z \in \mathcal{T}_{train}} loss(y_z, \hat{y}_z; \btheta).
\end{equation}
The notations and their descriptions are summarized in Table~\ref{tab:notation}. 

\begin{table}[t]
    \centering
    \caption{Notations and corresponding descriptions.}\label{tab:notation}
    \resizebox{\columnwidth}{!}{
        \begin{tabular}{c|l}
            \hline
            Notation & Description \\
            \hline
            $\mathcal{T}_{train}$, $\mathcal{T}_{test}$, $\mathcal{T}_{pool}$ & Training set, test set, retrieval pool. \\
            $s_z$,$s_t$ & The $z$-th sample, the target sample. \\
            $y_t, \hat{y}_t$ & The label and the predicted value of the target sample. \\
            $x_t$, $\mathbf{x}_t$ & The raw feature of $s_t$ and its embedding.\\
            $c_i^t$, $\mathbf{c}_i^t$ & The $i$-th feature in $x_t$ and its embedding.\\
            $\mathcal{R}(x_t)$ & The retrieved set of target sample $s_t$ using its feature $x_t$.\\
            $\bm{r}$, $\bm{l}$ & Aggregated feature and label representations of retrieved samples. \\
            $K$, $L$ & Number of retrieved samples, number of different labels.\\
            $d$ & Embedding size of each feature.\\
            \hline
            
        \end{tabular}
    }
\end{table}

\section{Methodology} \label{sec:method}
In this section, the details of RIM will be presented. 
We firstly give a big picture of our framework, then we elaborate on detailed descriptions about the two major modules of RIM.
Moreover, we will discuss the time complexity and some practical issues of RIM. Further discussions of RIM paradigm will also be included in this section.

\subsection{Overall Framework} 
\begin{figure}[h]
    \centering
    \includegraphics[width=0.9\columnwidth]{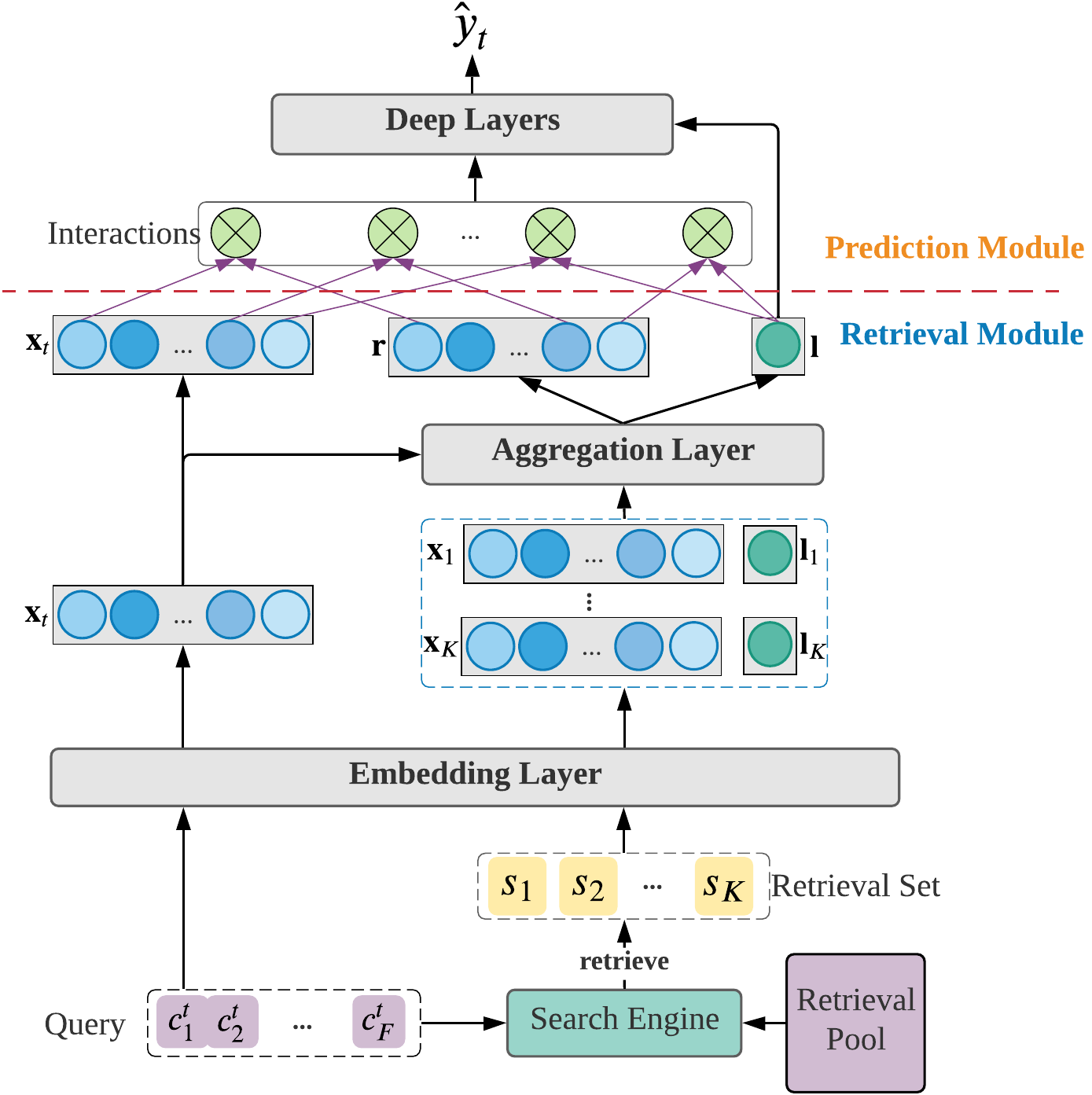}
    \caption{The overall illustration of RIM framework.}
    \label{fig:framework}
\end{figure}

As shown in Figure~\ref{fig:framework}, RIM consists of two major modules, i.e., retrieval module and prediction module. 

When there comes a request of prediction $s_t$, its features will be regarded as a query\footnote{In the following sections, "query" means the feature part of the target sample.}. The query is used to retrieve related samples from the retrieval pool $\mathcal{T}_{pool}$ through a search engine client.
After obtaining the corresponding embedding representations of retrieved samples and the target sample, there is an aggregation layer taking the query-aware order-invariant aggregation operation over representations of the retrieved samples, including not only feature part but also label part, since labels of the retrieved samples contain essential information. 

The prediction module takes the representations of aggregated retrieval samples and the target sample as the inputs. The features of them are combined and will be fed into the following feature interaction layer to explore the cross-row and cross-column interaction patterns. Finally, the following multi-layer perceptron (MLP) generates the output prediction.

\subsection{Retrieval Module} \label{sec:retrieval-module}
The retrieval module is responsible for retrieving data samples that have relations with the target sample. It utilizes search engine techniques to get the nearest data samples of the target in the raw sample space. This section will cover the details of data indexing, ranking function, and retrieval results aggregation.

\subsubsection{Indexing and Data Storage.}
\begin{figure}[t]
    \centering
    \includegraphics[width=1.0\columnwidth]{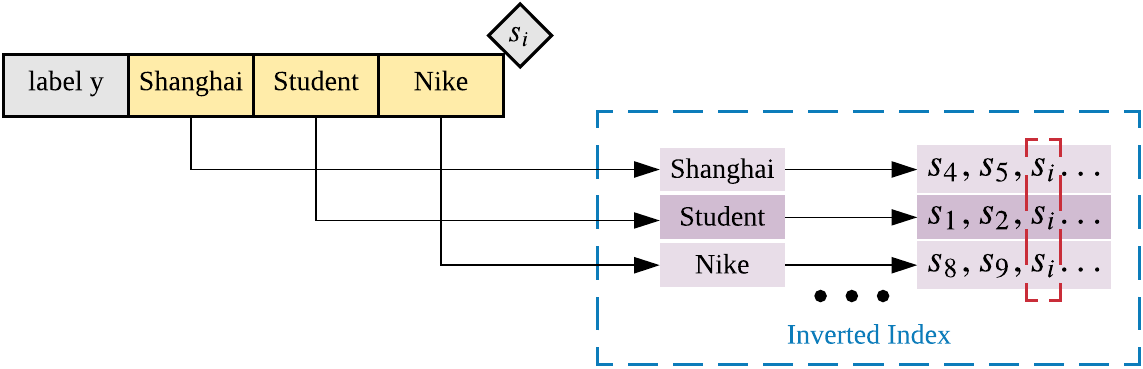}
    \caption{An illustration of the feature based inverted index.}
    \label{fig:i-index}
\end{figure}

We use the inverted index to store the data samples. Each data sample is regarded as a "document"\footnote{In the following sections, a "document" means the feature part of a sample in the retrieval pool $\mathcal{T}_{pool}$.}, and each feature of a sample is regarded as a "term"\footnote{In the following sections, "feature" and "term" are used interchangeably.}. As illustrated in Figure~\ref{fig:i-index}, the data sample $s_i$ has three feature fields, and $s_i$ is added to the three corresponding posting lists. Thus, for instance, the posting list of the term "Nike" consists of all the samples that has the feature "Nike". 

The numerical features should be discretized into categorical features because inverted index is more suitable for the discrete features that have enumerated values. As for the multi-value categorical features, in the indexing stage, we simply regard them as multiple terms and build the inverted index for each and every feature value in that field. For instance, if the "tag" feature of a sample has three values, "shopping", "mall" and "price comparison", then there will be three posting lists for all three feature values.

\subsubsection{Query.}
We use the features $x_t$ of the target sample $s_t$ as the query. As $x_t = \{c_i^t\}_{i=1}^F$, the logic of the query is formulated as

\begin{equation} \label{eq:query-form}
    c^t_1~OR~c^t_2~OR ...OR~c^t_F.
\end{equation}
By defining the target sample as the query, the search engine mechanism will retrieve the most similar samples with the target sample in the raw sample space. The query also represents a set of logic about how to select useful data from the dataset.

\subsubsection{Ranking Function.}
The samples that satisfy Eq.~\eqref{eq:query-form} will be ranked, and the top $K$ results will be retrieved.
We use commonly seen BM25 \cite{robertson1995okapi} as the ranking metric and the score between the query $x_t = \{c_i^t\}_{i=1}^F$ and a document $x_D = \{c_i^D\}_{i=1}^F$ is calculated as
\begin{equation}
    s = \sum_{i=1}^{F} \operatorname{IDF}(c_{i}^t) \cdot \frac{\text{TF}(c_{i}^t, x_D) \cdot(k_{1}+1)}{\text{TF}(c_{i}^t, x_D)+k_{1} \cdot(1-b+b \cdot \frac{|x_D|}{\text { avgdl }})}, 
\end{equation}
where $\text{TF}(c_{i}^t, x_D)$ is feature $c_i^t$'s term frequency in $x_D$.
If $c_i^t$ is a single value feature, then $\text{TF}(c_{i}^t, x_D)$ is either 1 or 0 depending on whether there is a match of $c_i^t$ in $x_D$.

If column $\mathcal{C}_i$ is a multi-value feature field, $c_i^t$ and $c_i^D$ both have multiple feature values. We denote them as $c_i^t = \{v_j\}_{j=1}^m$ and $c_i^D = \{u_j\}_{j=1}^n$. We use the Jaccard similarity as the term frequency which is calculated as
\begin{equation}
    \text{TF}(c_i^t, x_D) = \frac{|c_i^t \cap c_i^D|}{|c_i^t \cup c_i^D|}, c_i^D \in x_D,
\end{equation}
where $|\cdot|$ represent the size of a set.

The length of a document is defined as the number of feature fields, so all the documents have the same length $F$. Thus $\frac{|x_D|}{\text{avgdl}}=1$, where avgdl stands for the average document length. $k_1$ and $b$ are free parameters. 

IDF is defined as,
\begin{equation}
    \operatorname{IDF}(c_{i}^t)=\log \frac{\mathcal{N}-\mathcal{N}(c_{i}^t)+0.5}{\mathcal{N}(c_{i}^t)+0.5},
\end{equation}
where $\mathcal{N}$ is the total number of the documents and $\mathcal{N}(c_{i}^t)$ is the number of documents that contain the feature $c_i^t$. 
If $c_i^t = \{v_j\}_{j=1}^m$ is a multi-value feature, then the $\mathcal{N}(c_{i}^t)$ is calculated as 
\begin{equation}
    \mathcal{N}(c_{i}^t) = \frac{\sum_{j=1}^m\mathcal{N}(v_j)}{m}, \forall v_j\ \in c_{i}^t,
\end{equation}
which is the average number of documents that contains each feature value $v_j$.

The IDF term gives the common features less importance than rare features. It makes sense that a match on a rare feature implies stronger similarity signals compared to a commonly seen feature.

By ranking the candidate samples using BM25, we obtain the top $K$ samples as the retrieved set as shown in Figure~\ref{fig:framework}.

\subsubsection{Aggregation Layer.}
For the aggregation function, we use attention mechanism. The aggregated retrieved representation is calculated as
\begin{equation} \label{eq:atten-feats}
    \bm{r} = \sum_{k=1}^K \alpha_k \cdot \bm{x}_k, s_k \in \mathcal{R}(x_t),
\end{equation}
where $\mathcal{R}(x_t)$ is the retrieval set of target sample $s_t$ and $x_k$ is the feature part of the $k$-th retrieved sample $s_k$. The dense representation of $x_k$ is $\bm{x}_k \in R^{Fd \times 1}$ and $\bm{x}_k = \text{concat}([\bm{c}^k_1,...,\bm{c}^k_F])$ is the concatenation of the feature embeddings. $\alpha_k$ is the attention weight of the $k$-th retrieved sample which is defined as 
\begin{equation} 
    \alpha_k = \frac{\exp(\bm{x}_k^T \bm{W} \bm{x}_t)}{\sum_{j=1}^K \exp(\bm{x}_j^T \bm{W} \bm{x}_t)},
\end{equation}
where $\bm{W} \in R^{Fd \times Fd}$ is the attention layer parameter matrix.

The label information of the retrieved samples are also aggregated using the same attention weights as
\begin{equation}
    \bm{l} = \sum_{k=1}^K \alpha_k \cdot \bm{l}_k, s_k \in \mathcal{R}(x_t),
\end{equation}
where $\bm{l}_k$ is the label embedding that is looked up from an label embedding matrix $\mathcal{L} \in R^{L \times d}$ using corresponding label $y_k$ of retrieved sample $s_k$. If $y_k$ is a continuous value, it should be discretized first.

The feature embedding of target sample $\bm{x}_t$, the aggregated representations $\bm{r}$ and $\bm{l}$ are fed into the prediction module.

\subsection{Prediction Module}
The prediction module is responsible for calculating the final output of RIM based on the target sample and the retrieved samples.

\subsubsection{Interaction Layer.} \label{sec:interaction}
\begin{figure}[t]
    \centering
    \includegraphics[width=1.0\columnwidth]{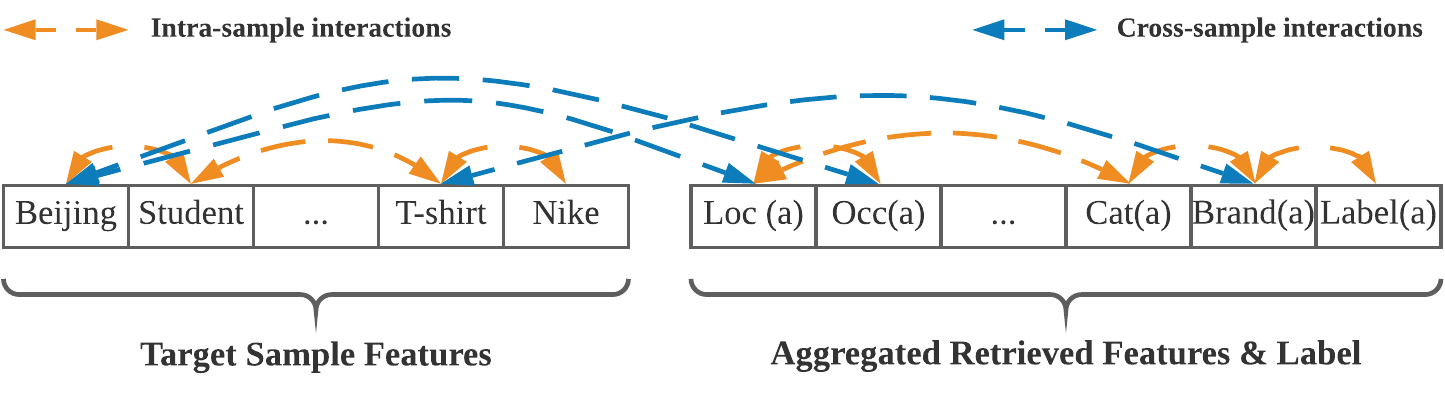}
    \caption{Illustration of intra-sample and cross-sample interactions. Loc (a), Occ (a), Cat (a), Brand (a) are the corresponding aggregated feature representations of the retrieved samples as described in Eq.~\eqref{eq:atten-feats}. (Loc=location, Occ=occupation, Cat=category).}
    \label{fig:inter_intuition}
\end{figure}

The major component of the prediction module is an interaction layer. 
As the representation of the retrieved samples is fed into the prediction module, it is beneficial to construct interacted features not only inside a sample representation (namely, intra-sample interactions) but between the target representation and retrieved representation (namely, cross-sample interactions), which is illustrated in Figure~\ref{fig:inter_intuition}. For example, the cross-sample interaction between the "occupation" feature of the target sample and that of retrieved samples (Occ(a)) will provide adequate collaborative filtering information. 

To model both the intra-sample and cross-sample interactions, the representations of the target sample and retrieved samples are concatenated as
\begin{equation}
    \bm{c}_{combine} = [\underbrace{\bm{c}^t_1, ..., \bm{c}^t_F}_{\bm{x}_t}, \underbrace{\bm{c}^a_1, ... \bm{c}^a_F}_{\bm{r}}, \bm{l}] = [\underbrace{\bm{e}_1, ..., \bm{e}_F}_{\bm{x}_t}, \underbrace{\bm{e}_{F+1}, ..., \bm{e}_{2F}}_{\bm{r}}, \bm{e}_{2F+1}],
\end{equation}
where we use $\bm{e}_p$ to simplify the notations.
By interacting all the features in $\bm{c}_{combine}$, both intra-sample and cross-sample interactions are modeled.
Thus the feature interactions are constructed as 
\begin{equation}
    \bm{e}_{inter} = [\text{inter}(\bm{e}_1, \bm{e}_2), \text{inter}(\bm{e}_1, \bm{e}_3), ..., \text{inter}(\bm{e}_{2F}, \bm{e}_{2F+1})],
\end{equation}
where $\text{inter}(\cdot)$ represents the interaction function. 
We could use different interaction functions here such as inner product \cite{rendle2010factorization, qu2016product, guo2017deepfm}, kernel product \cite{qu2018product} or micro-network \cite{qu2018product}.

 


\subsubsection{Output Layer.}
The output layer is fed with the concatenate vector of the former components as
\begin{equation}
    \textbf{inp} = \text{concat}([\bm{x}_t, \bm{r}, \bm{l}, \bm{e}_{inter}]).
\end{equation}
The output layer is an MLP equipped with a direct link of aggregated label embedding $\bm{l}$ at the final layer as shown in Figure~\ref{fig:framework}. The reason for the direct link of label representation is to strengthen the label information and to avoid too much interference with other features. The predicted value $\hat{y}_t$ is calculated as,
\begin{equation}
    \hat{y}_t = 
    \phi
    ([\sigma_{M}(\mathbf{W}_{M}(\ldots \sigma_{1}(\mathbf{W}_{1} \cdot \textbf{inp}+\mathbf{b}_{1}) \ldots)+\mathbf{b}_{M}), \bm{l}]),
\end{equation}
where $M$ is the number of layers, $\sigma_i$ is activation function, $W_i$ and $b_i$ are MLP parameters. The output of $\phi(\cdot)$ depends on the specific task. If it is a binary classification task, $\phi(\cdot)$ is a non-linear layer with sigmoid function. If it is a regression task, $\phi(\cdot)$ is a linear function.

\subsubsection{Loss function.}
For different tasks, we use different loss functions.
In our model, for binary classification and ranking tasks, we use cross-entropy loss. 
For regression task, we use MSE (mean square error) loss. Both loss functions are equipped with a typical L2-norm term with weight $\lambda$.
The form of loss function fully depends on the task.

\subsection{Practical Issues}
\subsubsection{Time Complexity.}
The most time-consuming part of RIM at inference time is the retrieval procedure, so in this section, we give the time complexity analysis of the retrieval module of RIM.
We use $|\mathcal{T}_{pool}|$ to denote the total number of samples inside $\mathcal{T}_{pool}$, and $V$ to denote the total number of unique features that have ever appeared in $\mathcal{T}_{pool}$. 
Then the average length of the posting lists in the inverted index is $\frac{|\mathcal{T}_{pool}|}{V}$. 
As the retrieval operation described in Section~\ref{sec:retrieval-module}, we first retrieve all the posting lists of features in $x_t$ which takes $\mathcal{O}(F)$ time, which is a constant.
The average number of retrieved samples is $F \cdot \frac{|\mathcal{T}_{pool}|}{V}$ and the scoring operation's complexity is linear to the number of retrieved samples as $\mathcal{O}(F \cdot \frac{|\mathcal{T}_{pool}|}{V})$.
So the entire time complexity of retrieval process is $\mathcal{O}(F) + \mathcal{O}(F \cdot \frac{|\mathcal{T}_{pool}|}{V}) = \mathcal{O}(F \cdot \frac{|\mathcal{T}_{pool}|}{V})$.

As for the time complexity of constructing the inverted index, it is $\mathcal{O}(F \cdot |\mathcal{T}_{pool}|)$ because each data sample will be appended to $F$ posting lists. This process could be done offline and only once. 

\subsubsection{Speed Up Approaches.}
It is normal to eliminate the feature fields with very few unique values, such as gender (only very few unique values). Such features appear in many samples, so the posting list is very long. We set this kind of feature fields as stop words to avoid a very long posting list. So the $\frac{|\mathcal{T}_{pool}|}{V}$ term could be limited.
We could also limit the size of $\mathcal{T}_{pool}$ by sampling to accelerate the retrieval process.

Another important way to speed up the retrieval process is to store the retrieval results instead of doing the retrieval procedure at every inference time. Instead, we do the retrieval process offline.
In the practice of recommender systems, for example, we could store the retrieval results for all the users with the corresponding recalled candidate items because the target samples consist of the user features and the candidate item features. 
It would be a trade-off between space and time complexity.



\section{Experiments} \label{sec:exp}
In this section, we present the experimental settings and corresponding results in detail. The experiments are conducted on three tasks, including CTR prediction, sequential top-$n$ recommendation, and rating prediction over 11 different datasets. For CTR prediction task, there are two groups of experiments involving sequential based models and feature interaction based models, respectively.
For each group of experiment, we compare RIM with the classic and state-of-the-art baselines over the suitable benchmark datasets, respectively. We have published the code of RIM \footnote{https://github.com/qinjr/RIM}.

This section starts with five research questions (RQs), and we use them to lead the following discussions.

\begin{itemize}[leftmargin=15pt]
  \item \textbf{RQ1:} Does RIM achieve the best performance compared to the baselines in all three kinds of tasks?
  \item \textbf{RQ2:} What is the performance of different retrieval approaches? Is the proposed retrieval method essential to the prediction performance?
  \item \textbf{RQ3:} How does the interaction function influence the prediction performance?
  \item \textbf{RQ4:} What is the influence of the retrieval set size $K$?
  \item \textbf{RQ5:} Is the label information of the retrieved samples essential to the performance?
\end{itemize}

\subsection{Experimental settings}\label{sec:exp-setting}
\subsubsection{Datasets} 
For different tasks, we use corresponding benchmark datasets. 
The details of the different datasets could be found in Appendix~\ref{apsec:datasets}. 
In total, we conduct experiments on four groups of datasets.

\begin{itemize}[leftmargin=15pt]
  \item \textbf{Group 1: Sequential based CTR prediction (Seq-CTR)}. For the sequential CTR prediction task that cares about how to model users' sequential behaviors, we choose three large-scale user behavior datasets from Alibaba and Ant Group. Tmall, Taobao and Alipay consist of user sequential behaviors collected on the corresponding e-commerce platforms.
  
  \item \textbf{Group 2: Feature interaction based CTR prediction (FI-CTR)}. For the feature interaction based baselines, we use Avazu and Criteo datasets, which are two commonly used CTR prediction benchmarks \cite{qu2018product,guo2017deepfm,liu2020autofis} by the feature interaction based CTR models.
  
  \item \textbf{Group 3: Sequential recommendation (Top-$n$ Ranking)}. We use Amazon Electronic (AZ-Elec), MovieLens-1m (ML-1m) and LastFM as benchmark datasets for sequential top-$n$ recommendation task.
  
  \item \textbf{Group 4: Regression (rating prediction, Reg)}. We choose Book-Crossing (BX), Amazon Cellphone and CCMR \cite{cao2016complete} of book rating, shopping rating and movie rating scenarios, respectively. 
\end{itemize}

\begin{figure}[t]
    \centering
    \includegraphics[width=1.0\columnwidth]{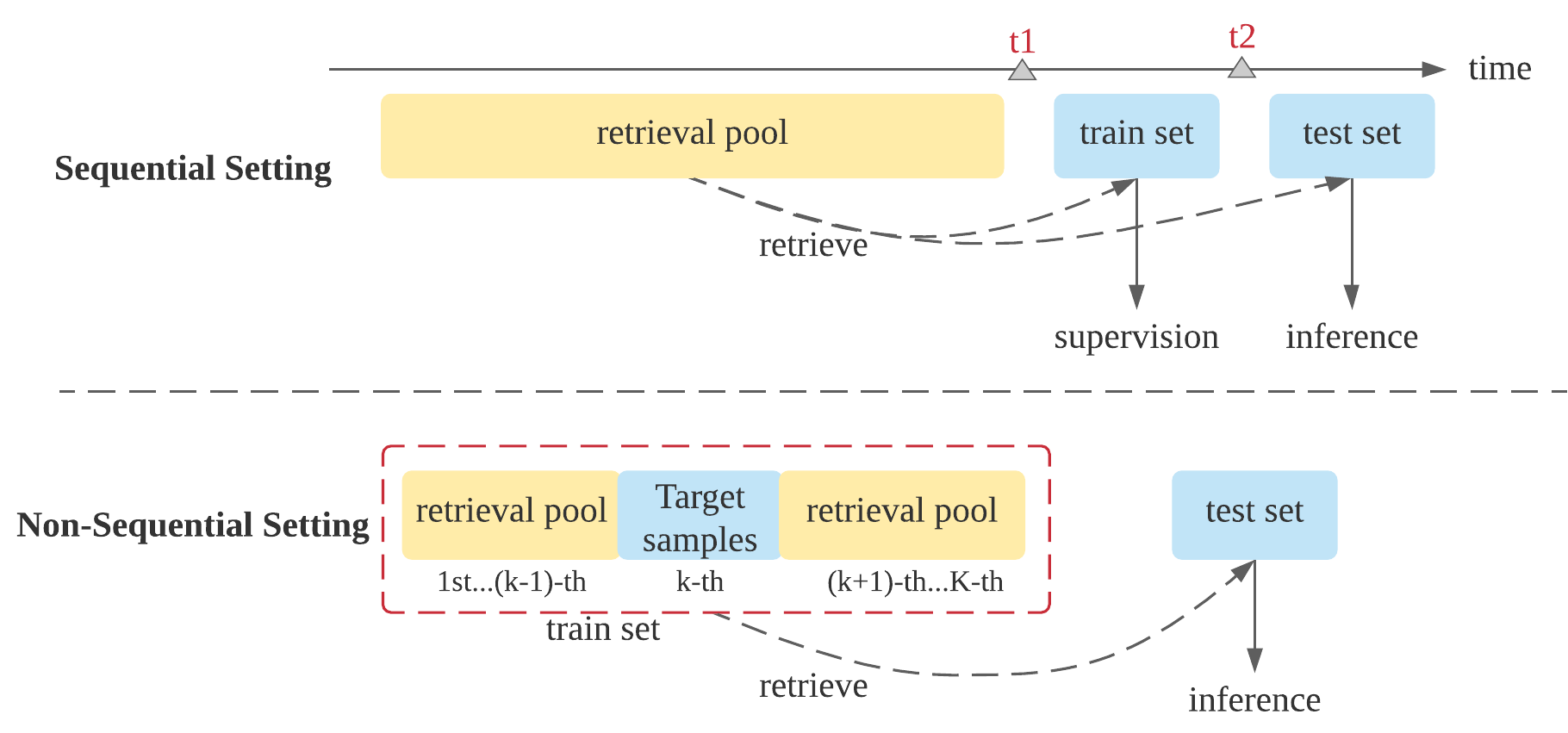}
    \caption{Two different data split scenarios. The upper part of the plot is for sequential setting while the lower plot is for non-sequential setting.}
    \label{fig:data_split}
\end{figure}

\textbf{Train \& Test \& Pool Splitting.}
There are two different splitting methods w.r.t different tasks as shown in Figure~\ref{fig:data_split}.

For the sequential modeling settings (Group 1 \& 3 experiments), we split the three parts of data using a global split time ($t_1$ and $t_2$ in Figure~\ref{fig:data_split}). If the timestamp of a sample is in the range of train/test/pool, it will be put into the corresponding set. 
It should be noticed that the splitting method is different from the widely used \textit{leave-one-out} strategy \cite{kang2018self, hidasi2015session, tang2018personalized} which splits the dataset using the relative sequential order of behaviors. Because using the global time to split the dataset instead of relative order is a better simulation of the real-world scenario.
The samples in the train and test set get the corresponding retrieved data from the pool. For the sequential baseline models, the sequential feature (e.g. user behaviors) is also generated from data of the retrieval pool.

For the non-sequential settings (non-sequential models or timestamp is not available, Group 2 \& 4), we use the train set as the retrieval pool for samples in the test set. For samples in the train set, we use a k-fold manner to retrieve neighbors for each sample. As shown in the lower part of Figure~\ref{fig:data_split}, for the target samples of the $k$-th fold, the other parts of the training set are used as the retrieval pool. For the baseline models, they just use the entire train set as a whole.
Our splitting method makes sure that all the baselines and RIM use the same amount of data and information in both settings.

\subsubsection{Evaluation Metrics}
For CTR prediction task (Group 1 \& 2), we choose the commonly used AUC and log loss to measure the performance, which reflects pairwise ranking performance and point-wise likelihood, respectively.
For the top-$n$ recommendation task (Group 3), we use hit ratio (HR@$K$), normalized discounted cumulative gains (NDCG@$K$), and mean reciprocal rank (MRR) to measure the list-wise ranking performance.
As for the regression task (Group 4), rooted mean squared error (RMSE) is utilized to test the rating prediction accuracy.

\subsubsection{Compared Baselines}

\begin{itemize}[leftmargin=15pt]
    \item \textbf{Group 1 (Seq-CTR)}. In this scenario, we compare RIM with six strong baselines of sequential behavior CTR models. HPMN \cite{ren2019lifelong} and MIMN \cite{pi2019practice} are memory network based models. DIN \cite{zhou2018deep} and DIEN \cite{zhou2019deep} are attention-based CTR models that capture user interests. SIM \cite{qi2020search} and UBR \cite{qin2020user} are retrieval based CTR models that retrieve useful user behaviors from lifelong user generated data.
    \item \textbf{Group 2 (FI-CTR)}. In this group of experiments, RIM is compared with the feature interaction based models. The selected baselines include traditional models like LR \cite{lee2012estimating}, GBDT \cite{chen2016xgboost} and FM-based models like FM \cite{rendle2010factorization}, AFM \cite{xiao2017attentional}, FFM \cite{juan2016field}. Deep feature interaction models like FNN \cite{zhang2016deep}, DeepFM \cite{guo2017deepfm}, IPNN \cite{qu2016product}, PIN \cite{qu2018product}, xDeepFM \cite{lian2018xdeepfm} and FGCNN \cite{liu2019feature} are selected as well.
    \item \textbf{Group 3 (Top-$n$ Ranking)}. For sequential recommendation (ranking) task, The selected baselines include PopRec which ranks items based on popularity, traditional MF-based recommendation models like BPR \cite{rendle2012bpr}, FPMC \cite{rendle2010factorizing}, TransRec \cite{he2017translation}.
    NARM \cite{li2017neural}, GRU4Rec \cite{hidasi2015session}, Caser \cite{tang2018personalized}, SASRec \cite{kang2018self}, SR-IEM \cite{pan2020rethinking} are the DNN based models that are proposed more recently.
    \item \textbf{Group 4 (Reg)}. In this group of experiments, we use rating prediction baselines such as FM \cite{rendle2010factorization}, AFM \cite{xiao2017attentional}, NFM \cite{he2017neural2}, NeuMF \cite{he2017neural} and IPNN \cite{qu2016product}.
\end{itemize}
The details of hyper parameters tuning could be found at Appendix~\ref{apsec:hyper-setting}. For each group of experiments, we have conducted significance test.

\subsection{Overall Performance Comparison: RQ1}
In total, we conduct four groups of experiments on 11 datasets. In this section, the overall experimental results are shown and discussed. We use inner product as the interaction function for the experiments in this section.
\subsubsection{Group 1 (Seq-CTR)} 
The results are listed in Table ~\ref{tab:ctr_seq}. From the results, we could find the following facts. 
(i) The performance of RIM is significantly better than the baselines. AUC values are improved by 1.81\%, 4.82\% and 0.67\% on three datasets compared to the best baseline, respectively.
(ii) We could find that retrieval-based sequential models (SIM, UBR) outperform other traditional models which only use the most recent consecutive behaviors. This fact shows retrieving the most relevant behavioral data instead of just using the most recent data is essential to sequential prediction performance.
(iii) RIM outperforms SIM and UBR significantly. Although UBR \& SIM retrieve the most relevant behaviors to predict, they are still in the single-row-multi-column framework. They just change how the sequential features are generated while RIM considers the cross-row patterns.

\begin{table*}
    \centering
    \caption{Performance comparison of CTR prediction task with sequential modeling baselines. Evaluation metrics are AUC and log-loss. Bold values are the best in each column, while the second best values are underlined. Rel.Impr means relative AUC improvement rate of RIM against each baseline. Improvements are statistically significant with $p < 0.01$.}
    \label{tab:ctr_seq}
    \begin{tabular}{c|ccc|ccc|ccc}
      \toprule
      \hline
      \multirow{2}{*}{ Model } & \multicolumn{3}{c|}{Tmall} & \multicolumn{3}{c|}{Taobao} & \multicolumn{3}{c}{Alipay} \\
      & AUC & LL & Rel.Impr & AUC & LL & Rel.Impr & AUC & LL & Rel.Impr \\
      \hline
      HPMN & 0.8526 & 0.4976 & 7.17\% & 0.7599 & 0.5911 & 12.68\% & 0.7681 & 0.5976 & 4.23\%\\
      MIMN & 0.8457 & 0.5008 & 8.05\% & 0.7533 & 0.6002 & 13.67\% & 0.7667 & 0.5998 & 4.42\%\\
      DIN & 0.8796 & \underline{0.4292} & 3.88\% & 0.7433 & 0.6086 & 15.20\% & 0.7647 & 0.6044 & 4.69\%\\
      DIEN & 0.8838 & 0.4445 & 3.39\% & 0.7506 & 0.6084 & 14.08\% & 0.7502 & 0.6151 & 6.71\%\\
      SIM & 0.8857 & 0.4520 & 3.17\% & 0.7825 & 0.5795 & 9.43\% & 0.7600 & 0.6089 & 5.34\%\\
      UBR & \underline{0.8975} & 0.4368 & 1.81\% & \underline{0.8169} & \underline{0.5432} & 4.82\% & \underline{0.7952} & \underline{0.5747} & 0.67\%\\
      \hline
      RIM & \textbf{0.9138} & \textbf{0.3804} & --- & \textbf{0.8563} & \textbf{0.4644} & --- & \textbf{0.8006} & \textbf{0.5615} & ---\\
      \hline
      \bottomrule
    \end{tabular}
\end{table*}

\subsubsection{Group 2 (FI-CTR)}
In this section, we compare RIM with the feature interaction based models. The results are shown in Table~\ref{tab:ctr_fi}.
From the table, we can observe that:
(i) RIM outperforms all the baselines in both datasets with 0.16\%-2.85\% and 0.04\%-2.88\% improvement rates, respectively.
(ii) PIN and xDeepFM use complex feature interaction functions, but RIM still outperforms them by simply using inner product interaction. Retrieving appropriate samples is more important than designing more complex interaction functions.
(iii) The improvement rate against FGCNN is limited because FGCNN generates a lot of new features using CNN. The number of parameters it uses is larger than RIM.

\begin{table}[!h]
    \centering
    \caption{Performance comparison of CTR prediction task with feature interaction based models. Improvements are statistically significant with $p < 0.01$.}
    \label{tab:ctr_fi}
    \resizebox{\columnwidth}{!} {
         \begin{tabular}{c|ccc|ccc}
         \toprule
          \hline
          \multirow{2}{*}{ Model } & \multicolumn{3}{c|}{Avazu} & \multicolumn{3}{c}{Criteo} \\
          & AUC & LL & Rel.Impr & AUC & LL & Rel.Impr\\
            \hline
            LR & 76.76\% & 0.3868 & 2.85\% & 78.00\% & 0.5631 & 2.88\% \\
            GBDT & 77.53\% & 0.3824 & 1.83\% & 78.62\% & 0.5560 & 2.07\% \\
            FM & 77.93\% & 0.3805 & 1.31\% & 79.09\% & 0.5500 & 1.47\% \\ 
            FFM & 78.31\% & 0.3781 & 0.82\% & 79.80\% & 0.5438 & 0.56\% \\ 
            AFM & 78.06\% & 0.3794 & 1.14\% & 79.13\% & 0.5517 & 1.42\% \\ 
            FNN & 78.30\% & 0.3778 & 0.83\% & 79.87\% & 0.5428 & 0.48\% \\ 
            DeepFM & 78.36\% & 0.3777 & 0.75\% & 79.91\% & 0.5423 & 0.43\% \\
            IPNN & 78.68\% & 0.3757 & 0.34\% & 80.13\% & 0.5399 & 0.15\% \\
            PIN & 78.72\% & 0.3755 & 0.29\% & 80.18\% & 0.5394 & 0.09\% \\
            xDeepFM & 78.55\% & 0.3766 & 0.51\% & 80.06\% & 0.5408 & 0.24\% \\ 
            FGCNN & \underline{78.82}\% & \underline{0.3747} & 0.16\% & \underline{80.22}\% & \underline{0.5389} & 0.04\% \\
            \hline
            RIM & \textbf{78.95}\% & \textbf{0.3741} & --- & \textbf{80.25}\% & \textbf{0.5387} & --- \\ 
            \hline
            \bottomrule
        \end{tabular}
    }
   
\end{table}

\subsubsection{Group 3 (Top-$n$ Ranking)}
For this group of experiments, we compare RIM with the sequential recommendation models. The ranking performance of the models is shown in Table~\ref{tab:rec-result}.
It should be noticed that the dataset splitting method is different from the common way \cite{kang2018self, tang2018personalized} that uses the relative order of user behaviors, but uses the global time, the performance of the baselines are thereby affected. 

From the table, we notice that RIM achieves significant improvements on recommendation performance against the baselines. As we only have one ground truth item for each ranked list, HR@$K$ is equivalent to Recall@$K$ and proportional to Precision@$K$, while MRR is equivalent to Mean Average Precision (MAP).
Different from Group 1 \& 2 experiments, the results of this group demonstrate RIM could perform well in list-wise ranking metrics which are important in recommendation tasks.

\begin{table*}
	\centering
	\caption{Performance comparison of sequential top-$n$ recommendation task in terms of HR, NDCG and MRR. Bold values are the best in each row, while the second best values are underlined. Improvements are statistically significant with $p < 0.01$.}\label{tab:rec-result}
	\resizebox{0.9\textwidth}{!}{
		\begin{tabular}{c|c|cccccccccc}
		    \toprule
			\hline
			Dataset & Metric & PopRec & BPR & FPMC & TransRec & NARM & GRU4Rec & Caser & SASRec & SR-IEM & RIM\\
            \hline
            \multirow{6}{*}{AZ-Elec} & HR@1 & 0.0645 & 0.0441 & 0.0435 & 0.0566 & 0.3049 & 0.3033 & 0.3041 & \underline{0.3051} & 0.3042 & \textbf{0.3248}\\
                                     & HR@5 & 0.2343 & 0.1642 & 0.1612 & 0.2057 & 0.5834 & 0.5807 & 0.5837 & \underline{0.5837} & 0.5801 & \textbf{0.6185}\\
                                     & HR@10 & 0.4160 & 0.2938 & 0.2895 & 0.3739 & 0.7259 & 0.7202 & 0.7264 & \underline{0.7269} & 0.7222 & \textbf{0.7366}\\
                                     & NDCG@5 & 0.1472 & 0.1027 & 0.1011 & 0.1294 & 0.4495 & 0.4473 & 0.4492 & \underline{0.4498} & 0.4477 & \textbf{0.4805}\\
                                     & NDCG@10 & 0.2055 & 0.1442 & 0.1423 & 0.1833 & 0.4956 & 0.4926 & 0.4954 & \underline{0.4961} & 0.4937 & \textbf{0.5188}\\
                                     & MRR & 0.1666 & 0.1237 & 0.1228 & 0.1512 & 0.4362 & 0.4338 & 0.4357 & \underline{0.4366} & 0.4349 & \textbf{0.4611}\\
			      \hline
            \hline
            \multirow{6}{*}{ML-1m} & HR@1 & 0.0285 & 0.0281 & 0.0261 & 0.0275 & 0.0337 & 0.0369 & 0.0371 & \underline{0.0392} & 0.0291 & \textbf{0.0645}\\
                                   & HR@5 & 0.1354 & 0.1312 & 0.1334 & 0.1375 & 0.1418 & 0.1395 & 0.1448 & \underline{0.1588} & 0.1331 & \textbf{0.2515}\\
                                   & HR@10 & 0.2669 & 0.2598 & 0.2577 & 0.2659 & 0.2631 & 0.2624 & 0.2597 & \underline{0.2709} & 0.2475 & \textbf{0.4014}\\
                                   & NDCG@5 & 0.0800 & 0.0779 & 0.0788 & 0.0808 & 0.0866 & 0.0872 & 0.0902 & \underline{0.0981} & 0.0802 & \textbf{0.1577}\\
                                   & NDCG@10 & 0.1222 & 0.1191 & 0.1184 & 0.1217 & 0.1254 & 0.1265 & 0.1271 & \underline{0.1341} & 0.1169 & \textbf{0.2059}\\
                                   & MRR & 0.1085 & 0.1053 & 0.1041 & 0.1078 & 0.1113 & 0.1135 & 0.1140 & \underline{0.1193} & 0.1058 & \textbf{0.1704}\\
			      \hline
            \hline
            \multirow{6}{*}{LastFM} & HR@1 & 0.0144 & 0.0191 & 0.0148 & 0.0563 & 0.0423 & \underline{0.0658} & 0.0542 & 0.0584 & 0.0584 & \textbf{0.0915}\\
                                    & HR@5 & 0.0947 & 0.1050 & 0.0733 & 0.1725 & 0.1394 & \underline{0.1785} & 0.1573 & 0.1729 & 0.1643 & \textbf{0.3468}\\
                                    & HR@10 & 0.2051 & 0.1965 & 0.1531 & 0.2628 & 0.2227 & \underline{0.2581} & 0.2336 & 0.2499 & 0.2495 & \textbf{0.5780}\\
                                    & NDCG@5 & 0.0522 & 0.0618 & 0.0432 & 0.1148 & 0.0916 & \underline{0.1229} & 0.1057 & 0.1163 & 0.1121 & \textbf{0.2165}\\
                                    & NDCG@10 & 0.0877 & 0.0910 & 0.0685 & 0.1441 & 0.1185 & \underline{0.1486} & 0.1303 & 0.1409 & 0.1394 & \textbf{0.2911}\\
                                    & MRR & 0.0822 & 0.0851 & 0.0694 & 0.1303 & 0.1083 & \underline{0.1362} & 0.1201 & 0.1289 & 0.1278 & \textbf{0.2210}\\
			\hline
			\bottomrule
		\end{tabular}
	}
	\vspace{-10pt}
\end{table*}

\subsubsection{Group 4 (Reg)}
For the regression task on explicit feedback, we choose the rating prediction problem. We compare our RIM framework with some well-known rating prediction models or models that can be used in this setting.
The results are shown in Table~\ref{tab:reg-result}.
RIM achieves 0.46\%, 1.87\%, and 0.09\% of relative improvement rates against the best baseline, respectively. 
These results could verify the efficacy of RIM in the regression task with explicit feedback, which indicates the universality of our framework.

\begin{table}[h]
    \centering
    \caption{Performance comparison of regression task in terms of RMSE (the lower, the better). 
    Improvements are statistically significant with $p < 0.01$.}
    \label{tab:reg-result}
    \resizebox{\columnwidth}{!}{
		\begin{tabular}{c|cc|cc|cc}
		  \toprule
		  \hline
          \multirow{2}{*}{ Model } & \multicolumn{2}{c|}{BX} & \multicolumn{2}{c|}{AZ-Cellphone} & \multicolumn{2}{c}{CCMR} \\
          & RMSE & Rel.Impr & RMSE & Rel.Impr & RMSE & Rel.Impr \\
          \hline
          FM                     & 2.6326    & 38.1\%       & 2.1408        & 36.2\%           & 0.9225     & 19.0\%       \\
          \hline
          AFM                    & 1.6936    & 3.33\%       & 1.4349        & 4.77\%           & 0.7594     & 1.36\%        \\
          \hline
          NFM                    & 1.6581    & 1.26\%       & 1.4251        & 4.11\%           &     0.7583    & 1.21\%        \\
          \hline
          NeuMF                  & \underline{1.6447}    & 0.46\%       & 1.3962        & 2.13\%           & 0.7537     & 0.61\%        \\
          \hline
          IPNN                   & 1.6464    & 0.56\%       & \underline{1.3925}        & 1.87\%           & \underline{0.7498}     & 0.09\%        \\
          \hline
          RIM                    & \textbf{1.6289}    & ---       & \textbf{1.3665}        & ---           & \textbf{0.7468}     & --- \\   
          \hline
          \bottomrule
        \end{tabular}
	}
\end{table}

\subsection{Ablation \& Hyper-parameter Study}
\subsubsection{Retrieval Mechanism: RQ2}
In this ablation experiment, we compare the performance of different retrieval mechanisms. Except for the retrieval mechanism that we present at Section~\ref{sec:retrieval-module}, random retrieval and filtered retrieval are used. Random retrieval randomly chooses some samples for the target. The filtered retrieval means that we first filter the retrieval pool with some specific constraints and then use the ranking mechanism mentioned in Section~\ref{sec:retrieval-module}. We use the user\_id feature to filter out the data that is not generated by the user herself.
The results are shown in Table~\ref{tab:rq2}. From the table, we could find that the retrieval mechanism of RIM is the best against other ways.
Random retrieval hurts the performance a lot because the retrieved samples do not have relations with the target sample.
The performance of using the filtered retrieval mechanism also drops, which demonstrates that retrieving from the entire sample space instead of the filtered samples is essential. We use different features to filter, it shows the similar trends.

\begin{table}[!h]
    \centering
    \caption{Performance of different retrieval mechanism.}
    \label{tab:rq2}
    \resizebox{\columnwidth}{!}{
        \begin{tabular}{c|cc|cc|cc}
          \toprule
          \hline
          \multirow{2}{*}{ Model } & \multicolumn{2}{c|}{Tmall} & \multicolumn{2}{c|}{Taobao} & \multicolumn{2}{c}{Alipay} \\
               & AUC    & LL     & AUC    & LL     & AUC    & LL     \\ \hline
            RIM-rand & 0.8366 & 0.5091 & 0.6034 & 0.6765 & 0.6067 & 0.6738 \\ 
            RIM-filter   & 0.9012 & 0.4007 & 0.8118 & 0.5289 & 0.7837 & \textbf{0.5537} \\ 
            RIM        & \textbf{0.9138} & \textbf{0.3804} & \textbf{0.8563} & \textbf{0.4644} & \textbf{0.8006} & 0.5615 \\ 
          \hline
          \bottomrule
        \end{tabular}
    }
\end{table}

\subsubsection{Interaction Functions: RQ3}
\begin{figure}[t]
    \centering
    \includegraphics[width=1.0\columnwidth]{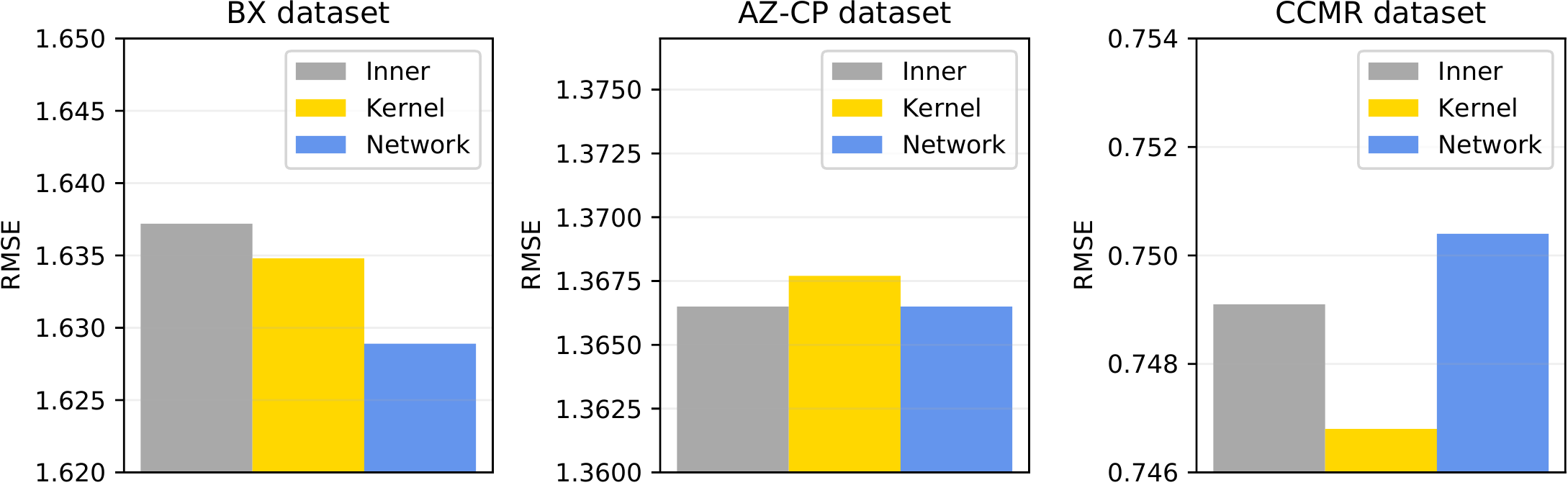}
    \caption{Performance of different interaction functions.} 
    \label{fig:rq3}
\end{figure}

We conduct experiments over different interaction functions described in Section~\ref{sec:interaction}. The results are shown in Figure~\ref{fig:rq3}.
The RMSE results on BX and CCMR datasets show that using kernel matrix or micro-networks could bring some improvements. However, on AZ-CP dataset, the inner product is nearly the same as the network product and better than the kernel product. 
The results indicate that using more complex interaction functions could be beneficial, but the inner product is good enough considering it does not bring more parameters. The inference time using inner product is faster than kernel product or micro-network by 11.8\% and 6.6\%, respectively. 

\subsubsection{Retrieval Set Size: RQ4}
\begin{figure}[!h]
    \centering
    \includegraphics[width=1.0\columnwidth]{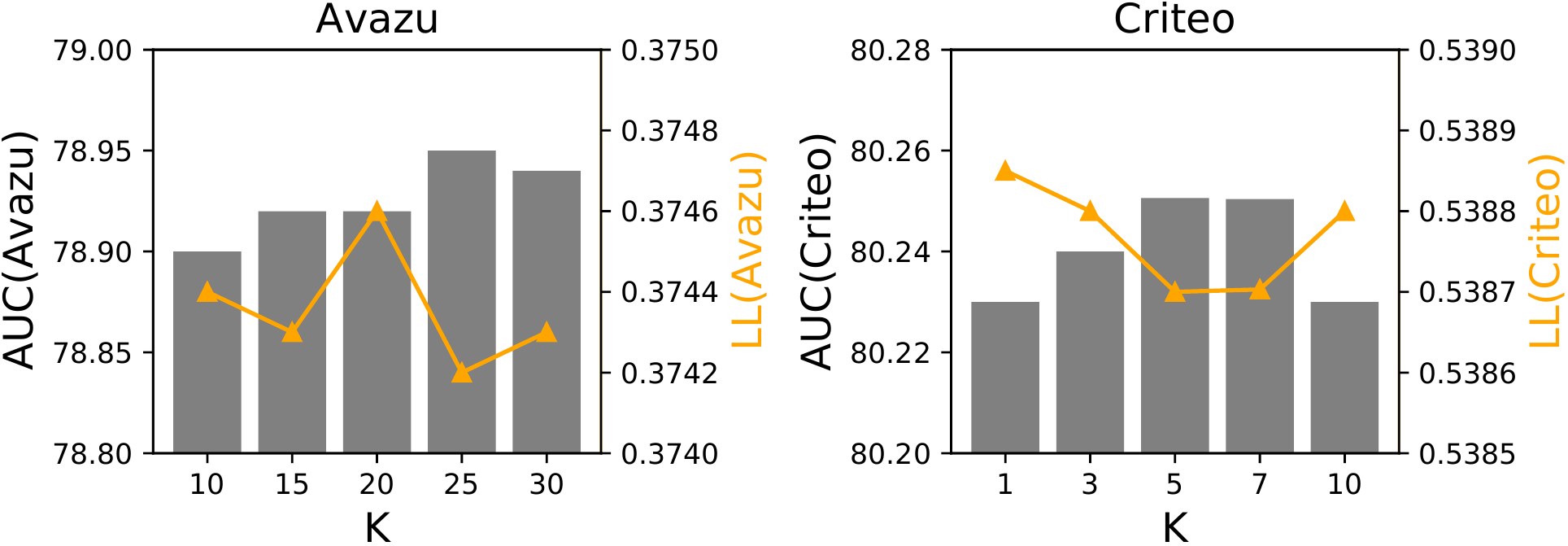}
    \caption{Performance of different retrieval sizes $K$.}
    \label{fig:rq4}
\end{figure}

In this section, we compare the performance of RIM w.r.t different retrieval sizes $K$ on Avazu and Criteo datasets.
From Figure~\ref{fig:rq4}, we can find that the fluctuation of AUC and log-loss is not very severe in terms of the absolute values.
However, there exists an optimal retrieval size for each dataset.
Naturally, too few retrieval samples may not be able to carry enough information to assist the prediction. Setting the size too large is not suitable either because too many retrieved samples will introduce much noise.

\subsubsection{Label information: RQ5}
In this section, we discuss the effectiveness of using label information of the retrieved samples. We remove the label part and only use the feature part of the retrieved samples. 
The results are listed in Table~\ref{tab:rq5}. The performance is much worse than RIM with label information, which demonstrates the necessity of incorporating label information.
As there are samples of different labels (e.g., positive and negative) in the retrieval pool, it is vital to consider the label of them because it is a strong indicator of the prediction. The model should be aware of the label attached to each retrieved sample. Thus it could have a better sense of the local information around the target sample in the sample space.

\begin{table}[!h]
    \centering
    \caption{Performance of RIM w/ and w/o label information.}
    \label{tab:rq5}
    \resizebox{\columnwidth}{!}{
    \begin{tabular}{c|cc|cc|cc}
        \toprule
        \hline
        \multirow{2}{*}{Model} & \multicolumn{2}{c|}{Tmall} & \multicolumn{2}{c|}{Taobao} & \multicolumn{2}{c}{Alipay} \\
                       & AUC    & LL     & AUC    & LL     & AUC    & LL     \\
       \hline
        RIM(w/o label) & 0.9093 & 0.3816 & 0.7921 & 0.5438 & 0.7635 & 0.5735 \\
        RIM            & \textbf{0.9138} & \textbf{0.3804} & \textbf{0.8563} & \textbf{0.4644} & \textbf{0.8006} & \textbf{0.5615} \\
        \hline
        \bottomrule
    \end{tabular}
    }
\end{table}

\section{Related Work} \label{sec:related}
\subsection{Models over Tabular Data}
The models over tabular data could be categorized into three parts. The first one is the traditional supervised learning models including LR, GBDT \cite{chen2016xgboost}, SVM et al. 
The second category is feature interaction based models. It originates from POLY2 \cite{chang2010training} and FM \cite{rendle2010factorization}. AFM \cite{xiao2017attentional} and FFM \cite{juan2016field} are variants of FM. After combined with DNNs structure \cite{cheng2016wide,shan2016deep}, various deep feature interaction models are proposed. DeepFM \cite{guo2017deepfm} and PNN \cite{qu2016product} uses MLP with the FM layer. PIN \cite{qu2018product} extends PNN by introducing a network-in-network structure to replace the inner product interaction function. FGCNN \cite{liu2019feature} incorporate CNNs to generate new features for interaction.
The third category is sequential models that focus on the sequential features. There are RNN-based models \cite{hidasi2015session,hidasi2017recurrent,beutel2018latent}, CNN-based models \cite{tang2018personalized} and Transformer-based models \cite{kang2018self}. Attention mechanism is also widely used in sequential models \cite{zhou2018deep,zhou2019deep,li2017neural}. To handle very long sequence, memory network based models \cite{ren2019lifelong,pi2019practice} are also proposed.
All of the above models use single row solely thus could be categorized as single-row-multi-column framework.

\textbf{Retrieval based models}.
UBR4CTR \cite{qin2020user} and SIM \cite{qi2020search} are two retrieval based models which retrieve useful behaviors instead of using the most recent ones. However, the retrieval module of them only affects how the sequential feature is generated. Therefore they are also categorized as single-row-multi-column.

\subsection{Neighbor-based Algorithms}
The basic idea of RIM is related to the $k$ nearest neighbor ($k$NN) algorithm, which also uses the neighbor samples to assist the inference of the target sample.
However, $k$NN is a passive learning algorithm without learnable parameters and uses only the labels of neighbor samples to generate predictions. 
RIM incorporates both features and labels of neighbor samples while models the complex interactions between the target sample and its neighbors.

\textbf{Neighbor-Augmented Neural Networks}.
There are neighbor-augmented deep models \cite{plotz2018neural,zhao2018retrieval,gu2018search} that share a similar motivation with RIM that use neighbor samples as an enhancement to model performance. 
However, these models ignore the interactions between target samples and its neighbors. Furthermore, RIM is the first neighbor-augmented deep model on tabular data while the existing models are designed for CV or NLP tasks.


\section{Conclusion And Future Work} \label{sec:conclusion}
In this paper, we propose the RIM framework for supervised learning tasks over tabular data. RIM makes full use of both cross-row and cross-column patterns of tabular data. It extends the traditional framework and achieves state-of-the-art performance on many different tasks of tabular data.

For the future work of this research, we plan to further investigate how to make RIM more efficient, and we are currently working on deploying it on a real-world advertisement system. Furthermore, exploring how to construct better index and retrieval functions is another interesting direction.

\section*{Acknowledgement}
This work is supported by ``New Generation of AI 2030'' Major Project (2018AAA0100900) and National Natural Science Foundation of China (62076161, 61772333, 61632017). The work is also sponsored by Huawei Innovation Research Programs. We thank MindSpore~\cite{mindspore} for the partial support of this work, which is a new deep learning computing framework.

\bibliographystyle{ACM-Reference-Format}
\bibliography{rst0078}
\clearpage
\appendix
\section{appendix}
\subsection{Details of the Datasets} \label{apsec:datasets}
\subsubsection{Dataset Preprocessing}
For RIM and non-sequential baselines, the datasets are processed into the tabular form, each row corresponds to a sample, and each column is a feature. For sequential baselines, the samples are sorted by timestamp.

\subsubsection{Retrieval Engine}
After the datasets are preprocessed, they are inserted into a retrieval engine using a comma-separated tokenizer. We use Elastic Search as the retrieval engine implementation, which is based on Apache Lucene.

\subsubsection{Datasets statistics}
In this section, we give the detailed dataset statistics and the URLs to get these datasets.

As shown in Table~\ref{tab:stat-datasets}, we use the 11 datasets of different scales and scenarios.
We count the number of users, items, samples, fields and categories. Number of categories is essentially the number of the unique feature values.

\begin{table}[!h]
  \centering
  \caption{The dataset statistics.}\label{tab:stat-datasets}
  \resizebox{\columnwidth}{!}{
    \begin{tabular}{c|c|c|c|c|c|c}
      \hline
      Task & Dataset & Users \# & Items \# & Samples \# & Field \# & Categories \#\\
      \hline\hline
      \multirow{3}{*}{Seq-CTR} & Tmall & 424,170 & 1,090,390 & 54,925,331 & 9 & 1,529,676 \\
      \cline{2-7}
      & Taobao & 987,994 & 4,162,024 & 100,150,807 & 4 & 5,159,462\\
      \cline{2-7}
      & Alipay & 498,308 & 2,200,291 & 35,179,371 & 6 & 3,327,205\\
      \hline
      \hline
      \multirow{2}{*}{FI-CTR} & Avazu & --- & --- & 40,428,967 & 24 & 645,195\\
      \cline{2-7}
      & Criteo & --- & --- & 99,616,043 & 39 & 1,178,909\\
      \hline
      \hline
      \multirow{3}{*}{Rec} & AZ-Elec & 728,719 & 160,052 & 6,739,591 & 3 & 890,156\\
      \cline{2-7}
      & ML-1m & 6,040 & 3,706 & 1,000,209 & 7 & 13,234 \\
      \cline{2-7}
      & LastFM & 991 & 1,076,238 & 18,993,371 & 5 & 1,184,264\\
      \hline
      \hline
      \multirow{3}{*}{Reg} & BX & 65,913 & 138,480 & 362,429 & 7 & 272,973\\
      \cline{2-7}
      & AZ-Cellphone & 6,211,701 & 589,534 & 10,063,256 & 3 & 6,802,474 \\
      \cline{2-7}
      & CCMR & 909,595 & 92,130 & 14,573,518 & 6 & 1,076,320\\
      \hline
    \end{tabular}
  }
\end{table}

For the ease of reproducing, the URLs to the raw datasets are given in Table~\ref{tab:datasets-url}. 
We would further provide the preprocessed datasets upon the acceptance of this paper.

\begin{table}[!h]
  \centering
  \caption{URLs to the datasets.}\label{tab:datasets-url}
  \resizebox{\columnwidth}{!}{
    \begin{tabular}{c|c}
      \hline
      Dataset & URL\\
      \hline
      Tmall & https://tianchi.aliyun.com/dataset/dataDetail?dataId=42\\
      Taobao & https://tianchi.aliyun.com/dataset/dataDetail?dataId=649\\
      Alipay & https://tianchi.aliyun.com/dataset/dataDetail?dataId=53\\
      Avazu & https://www.kaggle.com/c/avazu-ctr-prediction\\
      Criteo & https://labs.criteo.com/2013/12/download-terabyte-click-logs/\\
      AZ-Elec & https://jmcauley.ucsd.edu/data/amazon/\\
      ML-1m & https://grouplens.org/datasets/movielens/1m/\\
      LastFM & http://ocelma.net/MusicRecommendationDataset/lastfm-1K.html\\
      BX & http://www2.informatik.uni-freiburg.de/\textasciitilde cziegler/BX/\\
      AZ-Cellphone & https://jmcauley.ucsd.edu/data/amazon/\\
      CCMR & http://apex.sjtu.edu.cn/datasets/6\\
      \hline
    \end{tabular}
  }
\end{table}

\subsection{Hyperparameters Settings} \label{apsec:hyper-setting}
In this section, we give the detailed hyperparameter settings to reproduce the results in our paper.

\subsubsection{Tmall \& Taobao \& Alipay}
For these three datasets, we mainly tune the learning rate, l2\_norm, and batch size.
Learning rate is selected from $\{1 \times 10^{-3},5 \times 10^{-4},1 \times 10^{-4},5 \times 10^{-5}, 1 \times 10^{-5}\}$, l2\_norm is selected from $\{1 \times 10^{-4}, 5 \times 10^{-4}, 1 \times 10^{-5}, 5 \times 10^{-5}\}$, batch size is selected from $\{100, 200, 400\}$.

The tuning results are shown in Table~\ref{tab:hyper1}.

\begin{table}[!h]
    \resizebox{1\columnwidth}{!}{
    \begin{threeparttable}
    \caption{Hyper-parameter settings of Tmall, Taobao and Alipay.} \label{tab:hyper1}
    \label{tab:parameter}
    
    \begin{tabular}{c|l|l|l}
    \hline
    Params & Tmall & Taobao & Alipay \\
    \hline
    General & \begin{tabular}[c]{@{}l@{}}d=16; h=16\\ opt=Adam; net=[200,80,1]\end{tabular} & \begin{tabular}[c]{@{}l@{}}d=16; h=16\\ opt=Adam; net=[200,80,1]\end{tabular} & \begin{tabular}[c]{@{}l@{}}d=32; h=32\\ opt=Adam; net=[200,80,1]\end{tabular} \\
    \hline
    HPMN & \begin{tabular}[c]{@{}l@{}}lr=5e-4; l2=1e-5 \\ bs=100; up\_freq=[2,4,8]\end{tabular} & \begin{tabular}[c]{@{}l@{}}lr=1e-4; l2=5e-5 \\ bs=100; up\_freq=[2,4,8]\end{tabular} & \begin{tabular}[c]{@{}l@{}}lr=1e-4; l2=1e-5 \\ bs=200; up\_freq=[2,4]\end{tabular}\\
    \hline
    MIMN & \begin{tabular}[c]{@{}l@{}}lr=1e-4; l2=1e-5 \\ bs=100; mem\_layer=4\end{tabular} & \begin{tabular}[c]{@{}l@{}}lr=1e-4; l2=5e-5 \\ bs=100; mem\_layer=4\end{tabular} & \begin{tabular}[c]{@{}l@{}}lr=1e-4; l2=1e-4 \\ bs=100; mem\_layer=4\end{tabular}\\
    \hline
    DIN & \begin{tabular}[c]{@{}l@{}}lr=5e-4; l2=1e-4 \\ bs=100\end{tabular} & \begin{tabular}[c]{@{}l@{}}lr=5e-4; l2=5e-4 \\ bs=100\end{tabular} & \begin{tabular}[c]{@{}l@{}}lr=5e-4; l2=5e-4 \\ bs=100\end{tabular}\\
    \hline
    DIEN & \begin{tabular}[c]{@{}l@{}}lr=5e-4; l2=1e-4 \\ bs=200\end{tabular} & \begin{tabular}[c]{@{}l@{}}lr=5e-4; l2=1e-4 \\ bs=100\end{tabular} & \begin{tabular}[c]{@{}l@{}}lr=5e-4; l2=1e-4 \\ bs=100\end{tabular}\\
    \hline
    SIM & \begin{tabular}[c]{@{}l@{}}lr=1e-4; l2=5e-4 \\ bs=200; S=10\end{tabular} & \begin{tabular}[c]{@{}l@{}}lr=5e-4; l2=5e-4 \\ bs=200; S=10\end{tabular} & \begin{tabular}[c]{@{}l@{}}lr=5e-4; l2=5e-4 \\ bs=100; S=10\end{tabular}\\
    \hline
    UBR & \begin{tabular}[c]{@{}l@{}}lr\_pred=1e-4; lr\_retr \\l2=1e-5; bs=200\\ S=20\end{tabular} & \begin{tabular}[c]{@{}l@{}}lr\_pred=5e-4; lr\_retr \\l2=1e-5; bs=100\\ S=10\end{tabular} & \begin{tabular}[c]{@{}l@{}}lr\_pred=1e-4; lr\_retr \\l2=5e-5; bs=100\\ S=12\end{tabular}\\
    \hline
    RIM & \begin{tabular}[c]{@{}l@{}}lr=5e-4; l2=1e-4 \\ bs=100; S=10\end{tabular} & \begin{tabular}[c]{@{}l@{}}lr=1e-4; l2=5e-4 \\ bs=200; S=10\end{tabular} & \begin{tabular}[c]{@{}l@{}}lr=5e-4; l2=5e-4 \\ bs=100; S=10\end{tabular}\\
    \hline
    \end{tabular}

    \begin{tablenotes}
        \item[*] Note: bs= batch size, opt= optimizer, lr= learning rate, l2= $l_2$ regularisation weight, h= hidden size of GRU, d= embedding size, up\_freq= update frequency of HPMN, mem\_layer=number of memory layers (tracks) in MIMN, lr\_pred= learning rate of prediction module in UBR, lr\_retr= learning rate of retrieval module in UBR, net= MLP structure.
    \end{tablenotes}
    \end{threeparttable}
    }
\end{table}

\subsubsection{Avazu \& Criteo}
For this two dataset, the embedding size is searched from $\{20,40,60,80\}$, other hyperparameter settings are from \cite{qu2018product}. The tuned hyperparameters are shown in Table~\ref{tab:hyper2}.

\begin{table}[!h]
    \resizebox{0.9\columnwidth}{!}{
    \begin{threeparttable}
    \caption{Hyper-parameter settings of Criteo and Avazu.} \label{tab:hyper2}
    \label{tab:parameter}
    
    \begin{tabular}{c|l|l}
    \hline
    Params & Criteo & Avazu \\
    \hline
    General & \begin{tabular}[c]{@{}l@{}}bs=2000; lr=1e-3\\ opt=Adam; lr\_dc=0.9\end{tabular} & \begin{tabular}[c]{@{}l@{}}bs=2000; lr=1e-3\\ opt=Adam; lr\_dc=0.75\end{tabular}\\
    \hline
    LR & -- & -- \\
    \hline
    GBDT & \begin{tabular}[c]{@{}l@{}}depth=25;  \#tree=1300\end{tabular} & \begin{tabular}[c]{@{}l@{}}depth=18;  \#tree=1000\end{tabular}\\
    \hline
    \begin{tabular}[c]{@{}l@{}} FM,AFM\end{tabular} & \begin{tabular}[c]{@{}l@{}} n=20; t=0.01; h=32\\l2\_a=0.1; sub-net=[40,1]\end{tabular} & \begin{tabular}[c]{@{}l@{}} n=40; t=1; h=256;\\l2\_a=0; sub-net=[80,1]\end{tabular}\\
    \hline
    FFM & n=4 & n=4 \\
    \hline
    xDeepFM & \begin{tabular}[c]{@{}l@{}}n=20\\ net=[$400\times3$,1]\\CIN=[$100\times4$]\end{tabular} & \begin{tabular}[c]{@{}l@{}}n=40\\ net=[$700\times5$,1]\\ CIN=[$100\times2$]\end{tabular}\\
    \hline
    \begin{tabular}[c]{@{}l@{}} FNN,DeepFM,\\IPNN\end{tabular} & \begin{tabular}[c]{@{}l@{}}n=80\\ net=[$700\times5$,1]\\LN=true\end{tabular} & \begin{tabular}[c]{@{}l@{}}n=40\\ net=[$500\times5$,1]\\LN=true\end{tabular}\\ \hline
    PIN & \begin{tabular}[c]{@{}l@{}}n=40\\ net=[$700\times5$,1]\\ sub-net=[40,5]\\ LN=true\end{tabular} & \begin{tabular}[c]{@{}l@{}}n=40\\ net=[$500\times5$,1]\\ sub-net=[40,5]\\ LN=true\end{tabular}\\
    \hline
    FGCNN & \begin{tabular}[c]{@{}l@{}}k=20\\ conv=9*1\\ kernel=[38,40,42,44]\\ new=[3,3,3,3]\\ BN=T\\ net=[4096,2048,1]\end{tabular} & \begin{tabular}[c]{@{}l@{}}k=40\\ conv=7*1\\ kernel=[14,16,18,20]\\ new=[3,3,3,3]\\ BN=T\\ net=[4096,2048,\\ 1024,512,1]\end{tabular}\\
    \hline
    RIM & \begin{tabular}[c]{@{}l@{}}n=60\\ S=5 \\ LN=T\\ net=[$800\times5$,1]\end{tabular} & \begin{tabular}[c]{@{}l@{}}n=80\\ S=25 \\ LN=T\\ net=[$500\times5$,1]\end{tabular}\\
    \hline
    \end{tabular}

    \begin{tablenotes}
        \item[*] Note: bs= batch size, opt= optimizer, lr= learning rate, lr\_dc= learning rate decay factor, l2\_e= $l_2$ regularisation on embedding layer, t= softmax temperature, l2\_a= $l_2$ regularisation on attention network, h= attention network hidden size, n= embedding size, net= MLP structure, sub-net= micro network, LN= layer normalization, BN= batch normalization.
    \end{tablenotes}
    \end{threeparttable}
    }
\end{table}

\subsubsection{AZ-Elec \& ML-1m \& LastFM}
For these three datasets, we mainly tune the learning rate, l2\_norm and batch size.
Learning rate is selected from $\{1 \times 10^{-3},5 \times 10^{-4},1 \times 10^{-4},5 \times 10^{-5}, 1 \times 10^{-5}\}$, l2\_norm is selected from $\{1 \times 10^{-4}, 5 \times 10^{-4}, 1 \times 10^{-5}, 5 \times 10^{-5}\}$, batch size is selected from $\{100, 200, 300, 400, 500\}$.

The tuning results are shown in Table~\ref{tab:hyper3}.

\begin{table}[!h]
    \resizebox{1\columnwidth}{!}{
    \begin{threeparttable}
    \caption{Hyper-parameter settings of AZ-Elec, ML-1m and LastFM.} \label{tab:hyper3}
    \label{tab:parameter}
    
    \begin{tabular}{c|l|l|l}
    \hline
    Params & AZ-Elec & ML-1m & LastFM \\
    \hline
    General & \begin{tabular}[c]{@{}l@{}}d=16; h=16\\ opt=Adam; net=[200,80,1]\end{tabular} & \begin{tabular}[c]{@{}l@{}}d=16; h=16\\ opt=Adam; net=[200,80,1]\end{tabular} & \begin{tabular}[c]{@{}l@{}}d=16; h=16\\ opt=Adam; net=[200,80,1]\end{tabular} \\
    \hline
    PopRec & -- & -- & --\\
    \hline
    BPR & \begin{tabular}[c]{@{}l@{}}lr=5e-4; l2=1e-5 \\ bs=200;\end{tabular} & \begin{tabular}[c]{@{}l@{}}lr=1e-4; l2=5e-5 \\ bs=100\end{tabular} & \begin{tabular}[c]{@{}l@{}}lr=1e-3; l2=1e-4 \\ bs=200\end{tabular}\\
    \hline
    FPMC & \begin{tabular}[c]{@{}l@{}}lr=1e-3; l2=1e-4 \\ bs=100;\end{tabular} & \begin{tabular}[c]{@{}l@{}}lr=1e-4; l2=5e-5 \\ bs=100\end{tabular} & \begin{tabular}[c]{@{}l@{}}lr=5e-4; l2=5e-4 \\ bs=300\end{tabular}\\
    \hline
    TransRec & \begin{tabular}[c]{@{}l@{}}lr=1e-3; l2=1e-5 \\ bs=400;\end{tabular} & \begin{tabular}[c]{@{}l@{}}lr=1e-3; l2=5e-5 \\ bs=100\end{tabular} & \begin{tabular}[c]{@{}l@{}}lr=1e-3; l2=1e-4 \\ bs=200\end{tabular}\\
    \hline
    NARM & \begin{tabular}[c]{@{}l@{}}lr=1e-3; l2=1e-4 \\ bs=500;\end{tabular} & \begin{tabular}[c]{@{}l@{}}lr=1e-3; l2=1e-4 \\ bs=100\end{tabular} & \begin{tabular}[c]{@{}l@{}}lr=1e-3; l2=5e-4 \\ bs=200\end{tabular}\\
    \hline
    GRU4Rec & \begin{tabular}[c]{@{}l@{}}lr=1e-3; l2=1e-5 \\ bs=500;\end{tabular} & \begin{tabular}[c]{@{}l@{}}lr=1e-3; l2=1e-4 \\ bs=200\end{tabular} & \begin{tabular}[c]{@{}l@{}}lr=1e-3; l2=1e-4 \\ bs=200\end{tabular}\\
    \hline
    Caser & \begin{tabular}[c]{@{}l@{}}lr=1e-3; l2=1e-5 \\ bs=500;\end{tabular} & \begin{tabular}[c]{@{}l@{}}lr=1e-3; l2=1e-4 \\ bs=100\end{tabular} & \begin{tabular}[c]{@{}l@{}}lr=1e-3; l2=1e-4 \\ bs=200\end{tabular}\\
    \hline
    SASRec & \begin{tabular}[c]{@{}l@{}}lr=1e-3; l2=1e-5 \\ bs=500;\end{tabular} & \begin{tabular}[c]{@{}l@{}}lr=1e-3; l2=1e-5 \\ bs=200\end{tabular} & \begin{tabular}[c]{@{}l@{}}lr=1e-3; l2=1e-4 \\ bs=200\end{tabular}\\
    \hline
    SR-IEM & \begin{tabular}[c]{@{}l@{}}lr=1e-3; l2=1e-5 \\ bs=400;\end{tabular} & \begin{tabular}[c]{@{}l@{}}lr=1e-3; l2=1e-5 \\ bs=100\end{tabular} & \begin{tabular}[c]{@{}l@{}}lr=1e-3; l2=5e-4 \\ bs=200\end{tabular}\\
    \hline
    RIM & \begin{tabular}[c]{@{}l@{}}lr=1e-3; l2=1e-4 \\ bs=500; S=10\end{tabular} & \begin{tabular}[c]{@{}l@{}}lr=1e-3; l2=1e-4 \\ bs=100; S=10\end{tabular} & \begin{tabular}[c]{@{}l@{}}lr=1e-3; l2=1e-4 \\ bs=500; S=10\end{tabular}\\
    \hline
    \end{tabular}

    \begin{tablenotes}
        \item[*] Note: bs= batch size, opt= optimizer, lr= learning rate, l2= $l_2$ regularisation weight, h= hidden size of GRU, d= embedding size, net= MLP structure.
    \end{tablenotes}
    \end{threeparttable}
    }
\end{table}

\subsubsection{BX \& AZ-Cellphone \& CCMR}
For these three datasets, we mainly tune the learning rate, l2\_norm and batch size.
Learning rate is selected from $\{1\times 10^{-2}, 1 \times 10^{-3}, 5 \times 10^{-3}\}$, l2\_norm is selected from $\{1 \times 10^{-3}, 5 \times 10^{-3}, 1 \times 10^{-4}\}$, batch size is selected from $\{100, 200\}$.

The tuning results are shown in Table~\ref{tab:hyper4}.

\begin{table}[!h]
    \resizebox{1\columnwidth}{!}{
    \begin{threeparttable}
    \caption{Hyper-parameter settings of BX, AZ-Cellphone and CCMR.} \label{tab:hyper4}
    \label{tab:parameter}
    
    \begin{tabular}{c|l|l|l}
    \hline
    Params & BX & AZ-Cellphone & CCMR \\
    \hline
    General & \begin{tabular}[c]{@{}l@{}}d=16; opt=Adam\end{tabular} & \begin{tabular}[c]{@{}l@{}}d=16; opt=Adam\end{tabular} & \begin{tabular}[c]{@{}l@{}}d=16; opt=Adam\end{tabular} \\
    \hline
    FM & \begin{tabular}[c]{@{}l@{}}lr=1e-2; l2=5e-3 \\ bs=100\end{tabular} & \begin{tabular}[c]{@{}l@{}}lr=1e-4; l2=5e-5 \\ bs=100\end{tabular} & \begin{tabular}[c]{@{}l@{}}lr=1e-4; l2=1e-5 \\ bs=200\end{tabular}\\
    \hline
    AFM & \begin{tabular}[c]{@{}l@{}}lr=1e-2; l2=5e-3 \\ bs=200\end{tabular} & \begin{tabular}[c]{@{}l@{}}lr=5e-3; l2=1e-3 \\ bs=200\end{tabular} & \begin{tabular}[c]{@{}l@{}}lr=1e-3; l2=1e-4 \\ bs=200\end{tabular}\\
    \hline
    NFM & \begin{tabular}[c]{@{}l@{}}lr=1e-2; l2=1e-3 \\ bs=200; net=[40,20]\end{tabular} & \begin{tabular}[c]{@{}l@{}}lr=1e-3; l2=1e-4 \\ bs=200; net=[40,20]\end{tabular} & \begin{tabular}[c]{@{}l@{}}lr=1e-3; l2=1e-4 \\ bs=200; net=[40,20]\end{tabular}\\
    \hline
    NeuMF & \begin{tabular}[c]{@{}l@{}}lr=1e-2; l2=1e-4 \\ bs=100; net=[10,5]\end{tabular} & \begin{tabular}[c]{@{}l@{}}lr=1e-4; l2=5e-3 \\ bs=200; net=[10,5]\end{tabular} & \begin{tabular}[c]{@{}l@{}}lr=5e-3; l2=1e-4 \\ bs=100; net=[10,5]\end{tabular}\\
    \hline
    IPNN & \begin{tabular}[c]{@{}l@{}}lr=1e-2; l2=1e-4 \\ bs=200; net=[200,80,1]\end{tabular} & \begin{tabular}[c]{@{}l@{}}lr=5e-3; l2=5e-3 \\ bs=200; net=[200,80,1]\end{tabular} & \begin{tabular}[c]{@{}l@{}}lr=5e-3; l2=1e-4 \\ bs=100; net=[200,80,1]\end{tabular}\\
    \hline
    RIM & \begin{tabular}[c]{@{}l@{}}lr=1e-2; l2=1e-3 \\ bs=100; S=10\\net=[200,80,1]\end{tabular} & \begin{tabular}[c]{@{}l@{}}lr=1e-3; l2=1e-4 \\ bs=100; S=10\\net=[200,80,1]\end{tabular} & \begin{tabular}[c]{@{}l@{}}lr=5e-3; l2=1e-4 \\ bs=100; S=10\\net=[200,80,1]\end{tabular}\\
    \hline
    \end{tabular}

    \begin{tablenotes}
        \item[*] Note: bs= batch size, opt= optimizer, lr= learning rate, l2= $l_2$ regularisation weight, d= embedding size, net= MLP structure.
    \end{tablenotes}
    \end{threeparttable}
    }
\end{table}

\end{document}